\documentclass[twocolumn,10pt,superscriptaddress, longbibliography,amsmath,amssymb,aps,pra,floatfix]{revtex4-2}

\usepackage[T1]{fontenc}

\usepackage{graphicx}
\usepackage{tikz}
\usetikzlibrary{quantikz2}
\usepackage{mathtools}
\usepackage{amsthm}
\usepackage{comment}

\newcommand{\pr}{\mathrm{Pr}}
\newcommand{\ex}{\mathbb{E}}

\usepackage{color}

\usepackage[colorlinks=true,linkcolor=red,citecolor=blue, linktocpage=true,breaklinks=true]{hyperref}

\newcommand{\identity}{1\!\!1}

\begin{document}

\title{Palindromic Structure of Depth-Efficient Quantum Search Algorithms}

\author{Kun Zhang}
\email{kunzhang@nwu.edu.cn}
\affiliation{School of Physics, Northwest University, Xi'an 710127, China}
\affiliation{Shaanxi Key Laboratory for Theoretical Physics Frontiers, Xi'an 710127, China}
\affiliation{Peng Huanwu Center for Fundamental Theory, Xi'an 710127, China}
\affiliation{Fundamental Discipline Research Center for Quantum Science and Technology of Shaanxi Province, Xi'an 710127, China}

\author{Hai-Long Shi}
\email{hailong.shi@ino.cnr.it} 
\affiliation{INO-CNR and LENS, Largo Enrico Fermi 2, 50125 Firenze, Italy} 

\author{Xiao-Hui Wang}
\email{xhwang@nwu.edu.cn}
\affiliation{School of Physics, Northwest University, Xi'an 710127, China}
\affiliation{Shaanxi Key Laboratory for Theoretical Physics Frontiers, Xi'an 710127, China}
\affiliation{Peng Huanwu Center for Fundamental Theory, Xi'an 710127, China}
\affiliation{Fundamental Discipline Research Center for Quantum Science and Technology of Shaanxi Province, Xi'an 710127, China}

\author{Vladimir \surname{Korepin}}
\email{vladimir.korepin@stonybrook.edu}
\affiliation{C.N. Yang Institute for Theoretical Physics, Stony Brook University, New York 11794, USA}

\date{\today}

\begin{abstract}
Grover's algorithm is optimal in query complexity, but not necessarily in circuit depth. We formulate unstructured quantum search as a circuit-depth optimization problem and identify a critical depth ratio separating query optimality from depth optimality. The resulting depth-efficient search operators exhibit a palindromic structure, in which shallow diffusion-like operators symmetrically replace selected Grover diffusion layers while preserving efficient amplitude amplification. This structure yields a simple depth-efficiency criterion and an analytic expression for the minimal expected depth. Applying the framework to $X$-type mixers, local diffusion operators, and nested local diffusion operators, we obtain substantial depth reductions over standard Grover search. In particular, nested local constructions reduce the total circuit depth by about $40\%$ when the oracle and Grover diffusion operators have comparable depth. These results reveal the resource-dependent nature of quantum-search optimality and establish palindromic constructions as a systematic route to depth-efficient quantum search algorithms.
\end{abstract}

\maketitle

\emph{Introduction.---}
Search is a central computational primitive, underlying tasks from decision and optimization to cryptography and data analysis \cite{Knuth1998TAOCP3}.
For unstructured search over \(N=2^n\) items, Grover's algorithm provides a quadratic quantum speedup, using \(\mathcal O(\sqrt N)\) oracle queries
instead of \(\mathcal O(N)\) queries required classically
\cite{grover1996fast,Grover1997needle}. This query scaling is optimal in the standard oracle-query model \cite{BBBV97,BBHT98,zalka1999grover}.


However, optimality in oracle queries is not equivalent to optimality in circuit depth. Grover search alternates oracle calls with diffusion operators, so an implemented algorithm incurs the depth cost of both. Since the standard Grover diffusion operator is a global reflection on all qubits (equivalent to the $n$-qubit Toffoli gate), its realization can require substantial depth
\cite{Barenco1995ElementaryGates,Maslov2016RelativePhaseToffoli,Zindorf2025EfficientMCG}. This raises the central question of how to choose the operations between oracle calls so as to minimize the total circuit depth of quantum search algorithms.

Grover search already contains a structural freedom: oracle calls can be interleaved with different unitaries. This observation underlies amplitude amplification, fixed-point search, and generalized search algorithms
\cite{Grover1998AlmostAnyTransformation,Brassard2002AmplitudeAmplification,Grover2002Tradeoffs,Grover2005FixedPoint,Yoder2014FixedPointOptimal,Long2001ZeroFailure,Galindo2000FamilyGrover,Kato2005GroverLike,Tulsi2012GeneralFramework,Tulsi2015Faster,Jiang2017NearOptimal,Morales2018VariationallyLearningGrover},
and has recently motivated depth-oriented and hardware-efficient search
protocols
\cite{Zhang2020DepthOptimizationQuantumSearch,Wang2020ProspectGroverNISQ,Brianski2021IntroducingStructure,Liu2021HardwareEfficientQuantumSearch,Campos2024DepthScalingQAOA}.
Despite these advances, no general depth-optimization principle has yet been established that determines, within a unified framework, whether Grover's query-optimal construction is also depth-optimal.

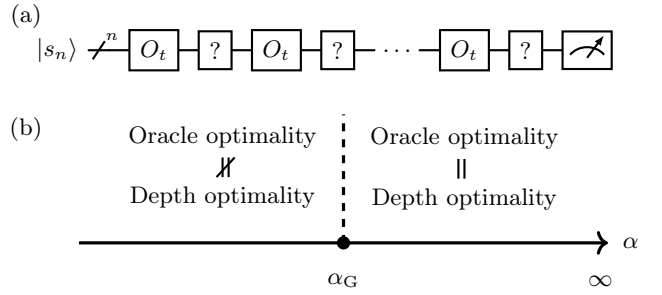
\begin{figure}[t]
$$
\begin{quantikz}[column sep=0.27cm]
\lstick{$|s_n\rangle$} & \qwbundle{n} & \gate{O_t} & \gate{?} & \gate{O_t} & \gate{?} & \ \ldots \ & \gate{O_t} & \gate{?} & \meter{}
\end{quantikz}
$$
\vspace{-1.5cm}
$$
\begin{tikzpicture}

    \draw (-3.5,0) edge [very thick,->] (3.5,0);
    \draw (0,0) edge [very thick,dashed] (0,1.7);

    \node [] at (0,-0.5) {$\alpha_\text{G}$};
    \fill (0,0) circle (2.5pt);
    \node [] at (3.4,-0.5) {$\infty$};

    \node [] at (3.8,0) {$\alpha$};

    \node [] at (1.6,1.38) {Oracle optimality};
    \draw [thick] (1.6,0.85) -- (1.6,1.11);
    \draw [thick] (1.52,0.85) -- (1.52,1.11);
    \node [] at (1.6,0.58) {Depth optimality};

    \node [] at (-1.6,1.4) {Oracle optimality};
    \draw [thick] (-1.6,0.87) -- (-1.6,1.13);
    \draw [thick] (-1.52,0.87) -- (-1.52,1.13);
    \draw [thick] (-1.44,1.13) -- (-1.68,0.87);
    \node [] at (-1.6,0.6) {Depth optimality};

    \node [] at (-4.2,1.5) {\small (b)};

    \node [] at (-4.2,3) {\small (a)};
\end{tikzpicture}\vspace{-0.5cm}
$$
\caption{(a) Depth-optimization formulation of quantum search, allowing the
oracle-independent interleaving operators to vary by layer.
(b) For $\alpha>\alpha_\text{G}$, Grover's query-optimal construction is also
depth optimal, whereas for $\alpha<\alpha_\text{G}$ shallower-than-Grover
search circuits exist. Here $\alpha$ and $\alpha_\text{G}$ are defined in
Eqs.~\eqref{eq:alpha} and \eqref{eq:alpha_G}.}
\label{fig_problem}
\end{figure}

In this Letter, we formulate quantum search as a circuit-depth optimization problem that allows different diffusion operators to be interleaved with oracle calls; see Fig.~\ref{fig_problem}. A central outcome of this optimization is the emergence of a palindromic structure: the optimized operators symmetrically replace selected Grover diffusion operators by shallower alternatives around a central Grover operator. This structure reduces circuit depth while preserving efficient amplitude amplification, and yields an analytically tractable expression for the minimal expected depth. We further identify a critical depth ratio \(\alpha_{\rm G}\) between the oracle and the standard Grover diffusion operator. Below this threshold, there exists a search algorithm that can outperform standard Grover search in depth. Explicit palindromic constructions based on different diffusion operators are demonstrated. In particular, nested local diffusion constructions can reduce the total circuit depth by about \(40\%\) relative to standard Grover search when the oracle and Grover diffusion operator have comparable depth. These results establish the palindromic structure as an emergent organizing principle for depth-efficient quantum search beyond the standard Grover algorithm.


\emph{Grover's algorithm.---}
The unstructured search problem seeks a marked state $t \in \{0,1\}^n$ among $N=2^n$ computational basis states. For simplicity, we assume a unique marked state. Classically, identifying the solution $t$ requires $\mathcal O(N)$ oracle queries. Grover's algorithm achieves a quadratic improvement through coherent amplitude amplification \cite{grover1996fast,Grover1997needle}. Starting from the uniform superposition $|s_n\rangle :=\sum_{j=0}^{N-1}|j\rangle/\sqrt N$, the oracle is implemented as the phase reflection $O_t := \identity - 2 |t\rangle\langle t|$, and the diffusion operator is $D_n := \identity - 2 |s_n\rangle\langle s_n|$. The Grover operator $G_n := D_n O_t$ is the product of two reflections and therefore acts as a rotation in the invariant two-dimensional subspace spanned by $|t\rangle$ and $|s_n\rangle$. Repeated applications of $G_n$ amplify the target-state amplitude and find the target state $t$ with near-unit probability after $\mathcal{O}(\sqrt{N})$ iterations, achieving optimal query complexity~\cite{BBBV97,BBHT98,zalka1999grover}.

The depth of $O_t$ is problem dependent, as it is determined by the reversible implementation of the underlying one-way function. Examples from cryptographic key search can be found in Refs.~\cite{Grassl2016ApplyingGroverAES,Almazrooie2018QuantumReversibleAES128,Jaques2020GroverAESLowMC,Langenberg2020ReducingAESCost,Liu2023ImprovedQuantumAES,Jang2025QuantumAnalysisAES}. By contrast, the diffusion operator $D_n$ is problem-independent and, up to single-qubit gates, equivalent to a generalized $n$-qubit Toffoli gate~\cite{nielsenQuantumComputationQuantum2010}. Its implementation on practical quantum computers typically requires circuit depth or gate count that scales with the number of qubits $n$~\cite{Zindorf2025EfficientMCG}. Thus, although $D_n$ is free in the oracle-query model, its circuit-depth cost is not generally negligible.

\emph{Depth-based formulation of quantum search.---}To describe a broad class of quantum search algorithms, we allow the oracle-independent operation inserted between oracle calls to vary from layer to layer. For a fixed number $k\in\mathbb N^+$ of oracle calls, we denote the set of allowed search operators as
\begin{equation}
\mathcal V_k[\mathcal U]
:=
\left\{
S_k = G_{U_k}G_{U_{k-1}}\cdots G_{U_1}
\;\middle|\;
U_l\in\mathcal U
\right\},
\end{equation}
where $G_U:=UO_t$ is the generalized Grover operator and $\mathcal U$ is an admissible set of oracle-independent interleaving operators. This formulation allows different layers to use different choices of $U$, and the standard Grover algorithm is recovered as the special case $\mathcal U=\{D_n\}$.

For a search operator \(S_k\in\mathcal V_k[\mathcal U]\) acting on the uniform superposition \(|s_n\rangle\), the single-run success probability is $\pr(S_k)=|\langle t|S_k|s_n\rangle|^2 $. Repetitions of the same circuit until success requires an average of \(1/\pr(S_k)\) runs, which gives the expected circuit depth
\begin{equation}
    \ex[S_k]:=
    \frac{\ell(S_k)}{\pr(S_k)},
\end{equation}
where \(\ell(S_k)\) denotes the circuit depth of one run of
\(S_k\). Given an admissible operator set
\(\mathcal U\), the depth-optimization problem is then
\begin{equation}
\label{eq:min_E}
    \ex_{\min}[\mathcal U]
    :=
    \min_{k\in\mathbb N^+}\;
    \inf_{S_k\in\mathcal V_k[\mathcal U]}
    \ex[S_k].
\end{equation}
The minimization is taken over the oracle count $k$ and over all admissible search operators $S_k\in\mathcal V_k[\mathcal U]$. We refer to $\ex_{\min}[\mathcal U]$ as the minimal expected depth (MED). For $\mathcal U=\{D_n\}$, this reduces to the MED of the standard Grover search.



We count the depth of oracle relative to the standard Grover diffusion operator and define
\begin{equation}
\label{eq:alpha}
    \alpha :=
    \frac{\ell(O_t)}{\ell(D_n)} ,
\end{equation}
so that $\ell(O_t)=\alpha\,\ell(D_n)$. Since Grover search is optimal in oracle-query complexity, when $\alpha\to\infty$, the oracle contribution dominates the circuit depth, so minimizing expected depth becomes equivalent, to leading order, to minimizing the expected number of oracle calls. Standard Grover search is therefore depth optimal in this limit. However, the limit $\alpha\to\infty$ is not practical. The central question is whether this equivalence survives at finite $\alpha$, where the depth of interleaving operators are no longer negligible. This motivates the critical depth ratio, defined for a given admissible set $\mathcal U$ as
\begin{equation}
\label{eq:alpha_G}
\alpha_\text{G}[\mathcal U]
:=
\sup\left\{
\alpha
\;\middle|\;
\ex_{\min}[\mathcal U]
<
\ex^\text{G}_{\min}
\right\},
\end{equation}
where $\ex^\text{G}_{\min}$ is the minimal expected depth of
standard Grover search. Thus, $\alpha_\text{G}[\mathcal U]$ marks the largest depth ratio below which the admissible family $\mathcal U$ can outperform Grover search in total circuit depth.

\emph{Depth-efficient palindromic search algorithms.---}
The optimization problem defined in Eq.~\eqref{eq:min_E} is combinatorial and challenging. Our first result is to identify a robust pattern via numerical optimizations, summarized as follows.  

\vspace{0.1cm}

\noindent\textbf{Result 1}. \emph{Depth-optimized search operators exhibit a palindromic structure, where the replacements are arranged symmetrically around the Grover operator $G_n$.} 

\vspace{0.1cm}

\noindent Details of the numerical optimizations are provided in the Supplemental Material (SM)~\cite{SM}. The simplest palindromic block is
\begin{equation}
\label{eq:palindromic_operator}
    P_U := G_U G_n G_{U^\dagger}.
\end{equation}
In particular, $G_{U^\dagger}=U^\dagger O_t$. This palindromic structure is robust against the choice of $U$. Grover's algorithm is a special case of $P_U$ with $U=D_n$. Therefore, we can view the palindromic construction as a generalization of Grover's algorithm.

The same idea extends naturally to nested palindromic blocks. For example, a two-level nesting gives $G_{U_2}G_{U_1}G_nG_{U_1^\dagger}G_{U_2^\dagger}$. More generally, for a prefix $G_{U_q}\cdots G_{U_1}$ with $q\in\mathbb N^+$, the nested palindromic block contains $2q+1$ oracle calls but only one Grover diffusion layer, replacing a
fraction $2q/(2q+1)$ of the Grover diffusion layers by shallower alternatives.

\begin{figure}[t]
\includegraphics[width=1\columnwidth]{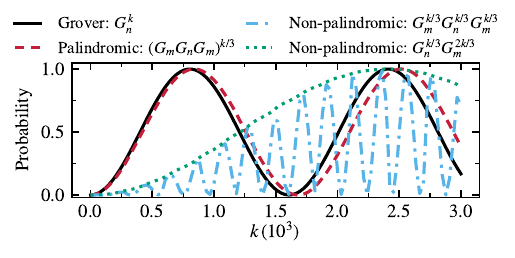}\vspace{-0.5cm}
\caption{Comparison of the success probabilities of Grover search $G_n^k$, the palindromic operator $(G_mG_nG_m)^{k/3}$, and two non-palindromic operators, $G_m^{k/3}G_n^{k/3}G_m^{k/3}$ and
$G_n^{k/3}G_m^{2k/3}$. Here $G_m=D_mO_t$, with $D_m$ defined in
Eq.~\eqref{eq:D_m}. For fixed total oracle count $k$, the three non-Grover operators have the same circuit depth. Parameters are $n=20$ and $m=5$.
}
\label{fig_compare}
\end{figure}

The key advantage of the palindromic structure is that it preserves efficient amplitude amplification while replacing most Grover diffusion layers by shallower operations. As shown in Fig.~\ref{fig_compare}, the palindromic operator maintains a
large success probability, whereas non-palindromic operators with the same circuit depth amplify the marked-state amplitude much less efficiently.

Formally we define the palindromic search algorithm as the circuit $P_U^kG_U$ applied to the initial state $|s_n\rangle$, followed by a measurement in the computational basis. This construction is motivated directly by the depth-based formulation of quantum search in Eq.~\eqref{eq:min_E}. The corresponding quantum circuit is \vspace{-0.3cm}
\[
\begin{quantikz}[row sep=0.1cm, column sep = 0.25cm]
\lstick{$|0\rangle^{\otimes n}$} & \qwbundle{n} 
& \gate{H^{\otimes n}} 
& \gate{G_U} 
& \gate{P_U} 
& \gate{P_U} 
& \ \ldots 
& \gate{P_U} 
& \meter{} \\
\end{quantikz}\vspace{-0.2cm}
\]
Here $H^{\otimes n}$ prepares the uniform initial state
$|s_n\rangle=H^{\otimes n}|0\rangle^{\otimes n}$
\cite{nielsenQuantumComputationQuantum2010}. For simplicity, we omit any work qubits required to implement the oracle $O_t$.

Our second main result is a simple depth-efficiency criterion for when a palindromic search algorithm outperforms standard Grover search in depth.

\vspace{0.1cm}

\noindent \textbf{Result 2}. \emph{The palindromic search algorithm has a smaller MED than Grover's algorithm if
\begin{equation}
\label{eq:optimal_criterion}
    \eta(P_U) < \eta(P_{D_n}),
    \qquad
    \eta(P_U):=\frac{\ell(P_U)}{\theta_U},
\end{equation}
where $\sin^2\theta_U=\left|\langle t|G_U|s_n\rangle\right|^2$.}

\vspace{0.1cm}

\noindent
The derivation is given in the End Matter. The quantity $\eta(P_U)$ is the depth
required per unit rotation angle toward the marked state. The guiding principle is to choose $U$ so as to generate a large rotation angle $\theta_U$ while keeping the depth $\ell(P_U)$ small. Note that $\eta(P_{D_n})$ is the depth efficiency of Grover's algorithm. 

Our third main result is a closed-form MED obtained from the depth optimization problem in Eq.~\eqref{eq:min_E} under the palindromic structure.

\vspace{0.1cm}

\noindent \textbf{Result 3}. \emph{
For an admissible set $\mathcal U=\{U,D_n\}$ satisfying Eq.~\eqref{eq:optimal_criterion}, the depth-efficient search with the palindromic structure has the MED
\begin{equation}
\label{eq:MED_palindromic}
    \mathbb{E}_{\min}[\mathcal U]
    \simeq
    0.69\,\eta(P_U).
\end{equation}}

\vspace{0cm}

\noindent
The derivation is given in the End Matter. The factor $0.69$ is universal and follows from optimizing the expected depth over the iteration number $k$. Thus Eq.~\eqref{eq:MED_palindromic} gives the analytic
MED associated with the palindromic optimum of Eq.~\eqref{eq:min_E}. The same argument extends directly to nested palindromic structures by replacing $P_U$
and $\theta_U$ with the corresponding nested block and rotation angle.


If near-unit success probability is desired, the same palindromic construction
also gives the corresponding optimal iteration number. The palindromic search algorithm finds the marked state with near-unit success
probability at
$k_{\max}=\left\lfloor \pi/(4\theta_U)-1/2 \right\rceil$, with circuit depth
$\ell_{\max}(P_U)\simeq \pi\eta(P_U)/4$. As a consistency
check, for $U=D_n$ one has $\theta_U\simeq 3/\sqrt{N}$ and
$\ell(P_{D_n})\simeq 3\ell(G_n)$, so the above formula recovers the standard
Grover scaling, namely $\ell_{\max}(P_{D_n})
    \simeq
    \pi\sqrt{N}\,\ell(G_n)/4$. 

Within the palindromic framework, the critical depth ratio
$\alpha_{\rm G}$ defined in Eq.~\eqref{eq:alpha_G} is obtained by solving
$\eta(P_U)=\eta(P_{D_n})$. For $\alpha<\alpha_{\rm G}$, we quantify the depth
advantage by the normalized depth reduction ratio (DRR)
\begin{equation}
\label{eq:DRR_palindromic}
    \mathrm{DRR}
    :=
    (1+\alpha)
    \left(
    1-\frac{\eta(P_U)}{\eta(P_{D_n})}
    \right).
\end{equation}
It quantifies the fractional total-depth reduction normalized by its largest
possible value. Indeed, in standard Grover search the oracle contribution alone
accounts for a fraction $\alpha/(1+\alpha)$ of the total depth, so replacing
diffusion layers can reduce the total depth by at most $1/(1+\alpha)$. The
prefactor $1+\alpha$ normalizes this upper bound to unity. 

Next, we present three examples on choosing $U$ from different classes of operators, and analyze their depth advantages relative to standard Grover search. These examples illustrate the design principle of applying $U$ that generates a large effective rotation angle $\theta_U$ while having a relatively small depth.

\emph{Optimization with $X$-type mixers.---}
The simplest shallow-depth choices for $U$ are single-qubit unitaries. Previous approaches replaced all Grover diffusion operators by single-qubit gates
\cite{Kato2005GroverLike,Jiang2017NearOptimal}, nearly doubling the oracle count and increasing the total depth when $\alpha>1$. By contrast, the palindromic construction replaces only two-thirds of the Grover diffusion layers, thereby yielding a net reduction in total circuit depth.

\begin{figure}[t]
\includegraphics[width=1\columnwidth]{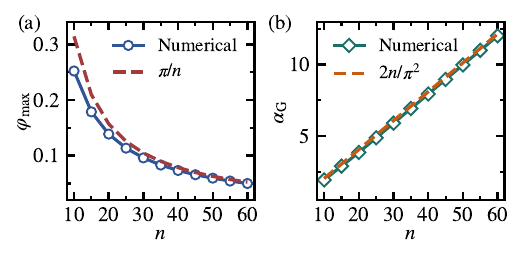}\vspace{-0.5cm}
\caption{(a) Optimal angle $\varphi$ for the unit-depth mixer $U=R_x(\varphi)^{\otimes n}$, obtained by minimizing
$\eta(P_\varphi)$. The large-$n$ prediction is $\varphi_{\max}=2\pi/n$.
(b) Critical depth ratio $\alpha_{\mathrm G}$ with large-$n$ prediction $\alpha_{\mathrm G}=2n/\pi^2$.
}
\label{fig_rx}
\end{figure}

Consider the unit-depth mixer $U=R_x(\varphi)^{\otimes n}$, with $R_x(\varphi)=e^{-i\varphi X/2}$. We write the corresponding
palindromic operator as $P_\varphi\equiv P_{R_x(\varphi)^{\otimes n}}$. To minimize the depth, the optimal angle $\varphi$ should maximize $\theta_U$, given by Eq. \eqref{eq:optimal_criterion}. In the large-$n$ limit this gives $\varphi_{\max}=2\pi/n$; see Fig.~\ref{fig_rx} and the SM \cite{SM}. The criterion \eqref{eq:optimal_criterion} then yields
\begin{equation}
    \alpha_\mathrm{G} = \frac{2n}{\pi^2} + \mathcal{O}(1).
\end{equation}
Moreover, $\mathrm{DRR}=2/3+\mathcal O(1/n)$, corresponding to a $33.3\%$ total-depth reduction at $\alpha=1$.
Thus even such a simple mixer separates query optimality from depth optimality at finite $\alpha$.

We next test whether correlated $X$-type mixers improve on the single-qubit construction. For $U=e^{-i\varphi \mathcal H_{XX}}$ with $\mathcal H_{XX}=\sum_j X_jX_{j+1}$, implementable in constant depth on platforms with native Ising-type interactions \cite{Andersen2025Thermalization}, we find only a small finite-size advantage. Its leading large-$n$ depth efficiency, and hence the scaling of $\alpha_{\mathrm G}$, is the same as for $R_x(\varphi)^{\otimes n}$. The global mixer $U=e^{-i\varphi X^{\otimes n}}$ is even less effective, since it couples each basis state only to its bitwise complement and does not outperform Grover search. Details are given in the SM \cite{SM}. The above examples suggest a simple design principle: $U$ should be a shallow surrogate of $D_n$, approximating its mean-inversion action on the search-relevant amplitudes so as to generate a large effective rotation angle $\theta_U$ while requiring a smaller depth than $D_n$.


\emph{Optimization with local diffusion operators.---}
As the Grover diffusion operator $D_n$ generates the largest rotation angle, a more intuitive construction is by choosing $U=D_m$, where
\begin{equation}
\label{eq:D_m}
    D_m = \mathbb I_{2^{n-m}} \otimes
    \left(\mathbb I_{2^m}-2|s_m\rangle\langle s_m|\right).
\end{equation}
It performs mean inversion only on an $m$-qubit subspace and thus serves as a
shallow surrogate of $D_n$. Since $\ell(D_m)<\ell(D_n)$ for $m<n$, $D_m$
interpolates between local diffusion and the Grover diffusion. Such operators were originally used in quantum partial search \cite{grover2004partial,korepin2005simple,Jiang2026ExactBounds}.

\begin{figure}[t]
\includegraphics[width=1\columnwidth]{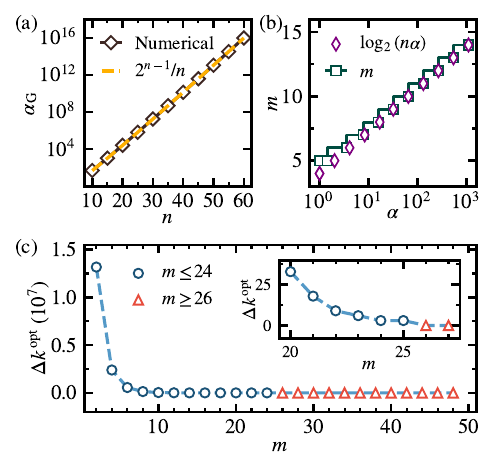}\vspace{-0.5cm}
\caption{(a) Critical depth ratio $\alpha_{\mathrm G}$ for the local-diffusion construction $U=D_m$, with large-$n$ prediction
$\alpha_{\mathrm G}=2^{n-1}/n$. (b) Optimal local size $m$ minimizing $\eta(P_m)$, with prediction $m=\lfloor\log_2(n\alpha)\rfloor$.
(c) Oracle-count difference $\Delta k^{\max}:=k_m^{\max}-k_n^{\max}$ between the palindromic construction and Grover search; the red triangle marks $\Delta k^{\max}=0$. The qubit number is set to $n=50$.}
\label{fig_triple}
\end{figure}

The palindromic operator $P_m\equiv P_{D_m}$ has an effective rotation angle $\theta_m$ satisfying
$\theta_n-\theta_m=\mathcal O(2^{-(m+n/2)})$, where
$\theta_n\equiv\theta_{D_n}$ is the Grover-diffusion angle. Thus $D_m$ rapidly approximates the rotation generated by $D_n$ while requiring smaller depth. Assume that $\ell(D_n)$ scales linearly with $n$ \cite{Zindorf2025EfficientMCG}, then it yields
\begin{equation}
    \alpha_\mathrm{G} = \frac{2^{\,n-1}}{n}+\mathcal O(1).
\end{equation}
The depth-advantage regime is therefore exponentially broad in $n$. For
$\alpha<\alpha_\mathrm{G}$, minimizing $\eta(P_m)$ gives
$m^\ast=\lfloor\log_2(n\alpha)\rfloor$ up to rounding, and
\begin{equation}
    \mathrm{DRR}
    =
    \frac{2}{3}
    \frac{n-\lfloor\log_2(\alpha n)\rfloor}{n}
    +\mathcal O(n^{-1}).
\end{equation}
Thus the DRR approaches $2/3$ at large $n$, but over an exponentially wider
range of depth ratios than in the single-qubit construction; see Fig.~\ref{fig_triple} and the SM \cite{SM}.

For circuits targeting near-unit success probability, the number of palindromic iterations is
\begin{equation}
    k_m^{\max}
    =
    k_n^{\max}
    +
    \frac{\pi}{9}2^{n/2-m}
    +
    \mathcal{O}\!\left(2^{-n/2}\right),
\end{equation}
where $k_n^{\max}$ denotes the corresponding Grover value. Hence, for
$m\gtrsim n/2$, replacing two thirds of the global diffusion layers by
local ones increases the optimal iteration number by less than one, in
agreement with Fig.~\ref{fig_triple}.

\emph{Optimization with nested local diffusion operators.---}
As a third construction, we use two local diffusion operators,
$\mathcal U=\{D_n,D_{m_1},D_{m_2}\}$ with $m_1>m_2$. Writing $G_m=D_mO_t$, numerical optimization over nested palindromic
sequences selects
\begin{equation} 
\label{eq:nested_P_U} 
\small 
P_{m_1,m_2,\tilde k} = \left(G_{m_2}G_{m_1}G_{m_2}\right)^{\tilde k} G_{m_2}G_nG_{m_2} \left(G_{m_2}G_{m_1}G_{m_2}\right)^{\tilde k}. \end{equation}
This structure gives the smallest MED among the nested structures tested. It contains $6\tilde k+3$ oracle calls, with $\tilde k$ the nesting depth,
and reduces to the previous local construction when
$\tilde k=0$.

\begin{figure}[t]
\includegraphics[width=1\columnwidth]{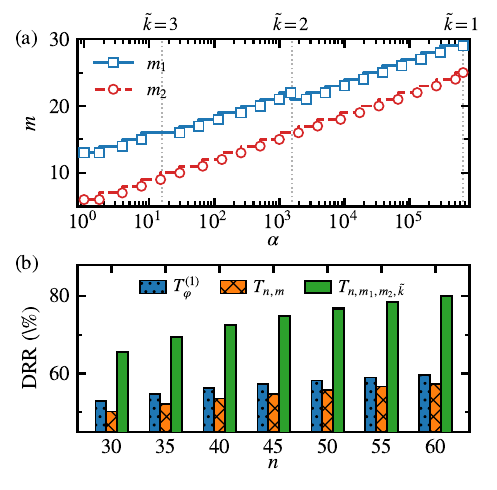}\vspace{-0.5cm}
\caption{(a) Optimal $m_1$, $m_2$, and $\tilde k$ for
$P_{m_1,m_2,\tilde k}$ in Eq.~\eqref{eq:nested_P_U} as functions of
$\alpha$ for $n=30$. Vertical dotted lines separate regions with different
optimal $\tilde k$. (b) DRR at $\alpha=1$ for three palindromic constructions: the single-qubit mixer $P_\varphi$, the local diffusion construction $P_m$, and the nested local-diffusion construction $P_{m_1,m_2,\tilde k}$, assuming linear scaling $\ell(D_n)\propto n$.
}
\label{fig_alpha_multi}
\end{figure}

The optimal parameters are strongly controlled by the depth ratio $\alpha$: $m_1$ and $m_2$ grow approximately logarithmically with $\alpha$, while smaller
$\alpha$ favors deeper nesting, namely larger optimal $\tilde k$; see Fig.~\ref{fig_alpha_multi}(a). This reflects the basic depth tradeoff: when oracle calls are cheaper, one can use more shallow local-diffusion layers to
replace global diffusion layers. As shown in Fig.~\ref{fig_alpha_multi}(b), the nested local-diffusion construction gives the largest DRR among the three
representative palindromic schemes. At $\alpha=1$, the DRR approaches $80\%$, corresponding to an actual depth reduction of about $40\%$ relative to Grover's algorithm.

These examples illustrate complementary routes to depth optimization within the palindromic framework. Single-qubit mixers offer a simple hardware-friendly implementation, whereas local and nested local diffusion operators better
approximate the mean-inversion action of $D_n$ and yield substantially larger depth reductions. Together, they establish palindromic constructions as a systematic route to depth-efficient quantum search beyond the standard Grover
search algorithm.

\emph{Conclusion and outlook.---} We have formulated unstructured quantum search as a circuit-depth optimization problem. It reveals that
Grover's oracle optimality need not imply circuit-depth optimality: the two
are separated by a critical depth ratio $\alpha_\text{G}$. We identify palindromic search as a
depth-efficient organizing principle and obtain the analytic MED in
Eq.~\eqref{eq:MED_palindromic}. Using single-qubit mixers, local diffusion
operators, and nested local diffusion operators, we demonstrate substantial
depth reductions over standard Grover search. Local diffusion extends the
depth-improvable regime exponentially with $n$, and nested local
diffusion achieves about a $40\%$ total-depth reduction at $\alpha=1$.
These results establish palindromic constructions as a systematic route to
depth-efficient quantum search and highlight the resource-dependent nature of
quantum-search optimality.

The palindromic framework also suggests hardware-aware implementations.
Because $U$ need not be a purely digital gate sequence, it can be realized
through hybrid analog-digital protocols \cite{ParraRodriguez2020DigitalAnalog}
or optimized directly at the pulse control level \cite{Alexander2020QiskitPulse}. Its symmetric, echo-like structure may also be compatible with pulse-level error-suppression techniques \cite{Viola1998DynamicalSuppression,Pokharel2024better}, suggesting a route toward depth-efficient and noise-resilient quantum search on realistic devices.

\emph{Acknowledgments.---} The authors thank Yun-Hao Shi, Kun Zhou, and Yan-Bo Jiang for helpful discussions. This work was supported by the NSFC (No. 12305028, No. 12275215), the Youth Innovation Team of Shaanxi Universities. KZ is supported by the China Postdoctoral Science Foundation under Grant Number 2025M773421, Shaanxi Province Postdoctoral Science Foundation under Grant Number 2025BSHYDZZ017, and Scientific Research Program Funded by Education Department of Shaanxi Provincial Government (Program No. 24JP186). HLS was supported by the Horizon Europe programme HORIZONCL4-2022-QUANTUM-02-SGA via Project No. 101113690 (PASQuanS2.1). VK is funded by the U.S. Department of Energy, Office of Science, National Quantum Information Science Research Centers, Co-Design Center for Quantum Advantage (C2QA) under Contract No. DE-SC0012704. 


%

\appendix
\section*{End Matter}
\setcounter{equation}{0}
\renewcommand{\theequation}{A\arabic{equation}}

\label{app:proof}

The palindromic search operator $P_U$ defined in
Eq.~\eqref{eq:palindromic_operator} preserves an effective
two-dimensional rotation structure, allowing coherent amplification of the
target-state amplitude despite the presence of multiple diffusion operators.
This allows the palindromic search algorithm to be analyzed within the
standard framework of amplitude amplification
\cite{Grover1998AlmostAnyTransformation,Brassard2002AmplitudeAmplification}.

Let $Q$ be an $n$-qubit unitary that prepares the state
\begin{equation}
    |s_Q\rangle := Q|0\rangle^{\otimes n}
    = \sum_{j=0}^{N-1} \alpha_j |j\rangle .
\end{equation}
Then $|s_Q\rangle$ can be decomposed into the target state $|t\rangle$ and the normalized non-target state $|nt\rangle$ as
\begin{equation}
    |s_Q\rangle
    =
    \sin\theta_Q |t\rangle
    +
    \cos\theta_Q |nt\rangle ,
\end{equation}
where $\sin\theta_Q=|\alpha_t|$ and $|nt\rangle$ is the normalized component
orthogonal to $|t\rangle$. The corresponding reflection operator is $D_Q := \mathbb I - 2|s_Q\rangle\langle s_Q|$.
Our convention differs from the more common reflection
$2|s_Q\rangle\langle s_Q|-\mathbb I$ only by an overall minus sign, which is
physically irrelevant. Therefore the generalized Grover operator
$G_Q=D_QO_t$ acts, up to a global phase, as a rotation in the
two-dimensional subspace spanned by $\{|t\rangle,|nt\rangle\}$. Ignoring the irrelevant global phase, after $k$ iterations one obtains
\begin{equation}
\label{eq:amplitude_amplification_k}
    G_Q^k |s_Q\rangle
    =
    \sin\!\left((2k+1)\theta_Q\right)|t\rangle
    +
    \cos\!\left((2k+1)\theta_Q\right)|nt\rangle .
\end{equation}
Accordingly, the success probability is
\begin{equation}
    \Pr(G_Q^k)
    =
    \sin^2\!\left((2k+1)\theta_Q\right).
\end{equation}
We now derive Results~2 and 3 by casting the palindromic search algorithm into
this amplitude-amplification form.

\begin{proof}[Derivation of Result 2]
The palindromic search operator $P_U$ defined in
Eq.~\eqref{eq:palindromic_operator} can be rewritten as
\begin{align}
P_U
&=
U O_t H^{\otimes n}
\bigl(\mathbb I-2|0\rangle^{\otimes n}\langle 0|^{\otimes n}\bigr)
H^{\otimes n} O_t U^\dagger O_t
\nonumber\\
&=
Q
\bigl(\mathbb I-2|0\rangle^{\otimes n}\langle 0|^{\otimes n}\bigr)
Q^\dagger O_t
\nonumber\\
&=
D_Q O_t,
\end{align}
where $Q:=UO_tH^{\otimes n}$. Thus the palindromic search algorithm is an amplitude-amplification process generated by $D_QO_t$.

Moreover, the initial state of the palindromic search algorithm satisfies
\begin{equation}
    G_U|s_n\rangle
    =
    UO_t|s_n\rangle
    =
    Q|0\rangle^{\otimes n}
    =
    |s_Q\rangle .
\end{equation}
Since $|\langle t|s_Q\rangle|
    =
    |\langle t|G_U|s_n\rangle|$, the angle $\theta_Q$ in the amplitude-amplification description is precisely
the angle $\theta_U$ defined in Eq.~\eqref{eq:optimal_criterion}. Therefore,
after $k$ palindromic iterations, the success probability is
\begin{equation}
\label{eq:palindromic_success_probability}
    \Pr(P_U^kG_U)
    =
    \sin^2\!\bigl((2k+1)\theta_U\bigr),
\end{equation}
up to the irrelevant global phase.

The total circuit depth after $k$ palindromic iterations is
\begin{equation}
    \ell_{\mathrm{tot}}(k)
    =
    \ell(G_U)+k\,\ell(P_U).
\end{equation}
Hence the MED within this palindromic construction is
\begin{equation}
\label{eq:appendix_MED_exact}
    \ex_{\min}(P_U)
    =
    \min_{k\in\mathbb N_0}
    \frac{\ell(G_U)+k\,\ell(P_U)}
    {\sin^2\!\bigl((2k+1)\theta_U\bigr)} .
\end{equation}
In the large-$N$ regime, $\theta_U\ll1$, and the initialization depth
$\ell(G_U)$ gives only a subleading additive correction compared with the
leading term of order $\ell(P_U)/\theta_U$. Thus
\begin{equation}
    \ex_{\min}(P_U)
    \simeq
    \min_{k\in\mathbb N_0}
    \frac{k\,\ell(P_U)}
    {\sin^2\!\bigl((2k+1)\theta_U\bigr)} .
\end{equation}
Treating $k$ as a continuous variable and setting
$x=(2k+1)\theta_U$, the minimization condition is $\tan x = 2x$. Its first positive solution is $x\simeq1.1656$, giving
\begin{equation}
    k_{\min}
    =
    \left\lfloor
    \frac{0.5828}{\theta_U}-\frac{1}{2}
    \right\rceil ,
\end{equation}
where $\lfloor\cdot\rceil$ denotes the nearest integer. Substitution gives
\begin{equation}
\label{eq:MED_palindromic_End_matter}
    \ex_{\min}(P_U)
    \simeq
    0.69\,\frac{\ell(P_U)}{\theta_U}
    =
    0.69\,\eta(P_U),
\end{equation}
where $\eta(P_U)=\ell(P_U)/\theta_U$.

The Grover construction is recovered by taking $U=D_n$, for which the same
asymptotic expression gives
\begin{equation}
    \ex_{\min}(P_{D_n})
    \simeq
    0.69\,\eta(P_{D_n}) .
\end{equation}
Therefore, to leading order in the large-$N$ regime, the inequality $\ex_{\min}(P_U)<\ex_{\min}(P_{D_n})$ gives $\eta(P_U)<\eta(P_{D_n})$. Thus the palindromic search algorithm has a smaller MED than Grover search
whenever $\eta(P_U)<\eta(P_{D_n})$, proving Result~2.
\end{proof}

\begin{proof}[Derivation of Result 3]
Equation~\eqref{eq:MED_palindromic_End_matter} gives the leading asymptotic MED for a palindromic search algorithm generated by a fixed block $P_U$. For the admissible set $\mathcal U=\{U,D_n\}$, the corresponding Grover
palindromic block is $P_{D_n}$. Under the condition $\eta(P_U)<\eta(P_{D_n})$, the block $P_U$ yields a smaller leading MED than the Grover block.
Therefore, within the palindromic construction identified in Result 1, the
depth-optimized search algorithm for $\mathcal U=\{U,D_n\}$ is generated by
$P_U$, and its MED is
\begin{equation}
    \ex_{\min}[\mathcal U]
    \simeq
    0.69\,\eta(P_U).
\end{equation}
This demonstrates Result~3 within the palindromic construction.
\end{proof}

The success probability given by Eq. \eqref{eq:palindromic_success_probability} is maximized when $(2k+1)\theta_U$ is closest to $\pi/2$. Therefore the
iteration number reaching the first maximum is
\begin{equation}
    k_{\max}
    =
    \left\lfloor
    \frac{\pi}{4\theta_U}
    -
    \frac{1}{2}
    \right\rceil .
\end{equation}
The rounding error is
at most $\theta_U$, and hence the corresponding success probability is
$1-\mathcal O(\theta_U^2)$ in the regime $\theta_U\ll1$. The total circuit depth at this iteration number is
\begin{equation}
    \ell_{\max}(P_U)
    =
    \ell(G_U)+k_{\max}\ell(P_U).
\end{equation}
Using $k_{\max}=\pi/(4\theta_U)+\mathcal O(1)$, we obtain
\begin{equation}
    \ell_{\max}(P_U)
    =
    \frac{\pi}{4}\frac{\ell(P_U)}{\theta_U}
    +
    \mathcal O\!\left(\ell(P_U)\right).
\end{equation}
In the asymptotic regime $\theta_U\ll1$, the additive
$\mathcal O(\ell(P_U))$ term, including the initialization depth
$\ell(G_U)$ and the integer-rounding correction, is subleading. Therefore
\begin{equation}
    \ell_{\max}(P_U)
    \simeq
    \frac{\pi}{4}\,\eta(P_U).
\end{equation}

\end{document}